# Metadisorder for designer light in random-walk systems

Sunkyu Yu, Xianji Piao, Jiho Hong, and Namkyoo Park*

*Photonic Systems Laboratory, Department of Electrical and Computer Engineering, Seoul National University, Seoul 08826, Korea*

*E-mail address for correspondence: nkpark@snu.ac.kr*

Disorder plays a critical role in signal transport, by controlling the correlation of systems[1]. In wave physics, disordered potentials suppress wave transport[2-4] due to their localized eigenstates from random-walk scattering. Although the variation of localization with tunable disorder[4-8] has been intensively studied as a bridge between ordered and disordered media, the general trend of disorder-enhanced localization[5] has remained unchanged, failing in envisaging the existence of delocalization in highly-disordered potentials. Here, we propose the concept of 'metadisorder': tunable random-walk systems having a designed eigenstate with unnatural localization. We demonstrate that one of the eigenstates in a randomly-coupled system can always be arbitrarily molded, regardless of the degree of disorder, by adjusting the self-energy of each element. Ordered waves are then achieved in highly-disordered systems, including planewaves and globally-collective resonances. We also devise counterintuitive functionalities in disordered systems, such as 'small-world-like'[1] transport from non-Anderson-type localization, phase-conserving disorder, and phase-controlled beam steering.

Network modeling has provided an intuitive picture for understanding various complex systems in nature and society. In these system networks, a 'disorder' is a crucial factor in signal transports over the system. For example, in the pioneering work[1] in graph theory, D. J. Watts discovered 'small-world disordered networks' with randomly rewired connections, which can model various elemental systems in physics, biology, and sociology: seismic networks[9], *C. elegans* neurons[10], brain connectome[11], and affinity groups in social networks[1].

The role of the disorder is also evident in wave physics, especially when compared to the 'order' in potential energy, such as periodic or quasiperiodic potentials. Periodic potentials allow ballistic transport through extended Bloch eigenstates[12], whereas broken correlation in disordered potentials leads to Anderson-localized eigenstates[2-4], which significantly suppress wave transport. Although serious attempts have been made[4-8] to fill the gap between order and disorder in wave systems, the increase of disorder has only led to a monotonous change from extended to localized eigenstates[5], prohibiting the existence of eigenstates with counterintuitive forms, for example, completely delocalized eigenstates in highly disordered potentials. Thus, a clear distinction has been maintained between the applications of order and disorder in wave systems, in accordance with the contrast between their eigenstates: transporting devices using ordered potentials[12,13] and focusing devices using disordered potentials[14-16]. The design of nontrivial waves, such as globally collective and scattering-free propagations, in disordered systems remains a challenge. The question thus naturally arises as to whether waves analogous to those in an ordered system can also be achieved in a highly disordered system, that is, the disorder being 'analogous' to the order and, if so, how this can be achieved.

In this paper, we demonstrate the existence of globally collective and delocalized waves in disordered systems. Unnatural phenomena distinctive from classical random-walk systems are demonstrated in these 'artificial' disordered optical potentials of controlled self-energy distributions: perfect planewaves without any scattering or phase distortion, designer guided-waves, globally collective resonances, conservative waves in complex potentials, and the invisible disorder of phase conservation, all of which provide ordered waves with disorder-like energy bands. By creating artificial disorder to achieve non-Anderson-type localization, we also reveal the counterintuitive relationship between eigenstate localization and wave transport, analogous to the 'small-world network'[1], which allows transport with clustering. Our method, paving the way toward disorder-robust small-world optics, is also applied to functional wave devices in highly disordered potentials, including tunable focusing, in-phase spatial oscillation, parity converters, and the point-source excitation of planewaves.

We start with random-walk systems composed of weakly coupled optical elements, such as waveguides[12,17] or resonators[18,19]. Based on discrete models of coupled-mode theory (CMT)[17] or tight-binding (TB) analysis[19], an $N$-body system is governed by the eigenvalue equation $H\psi = \gamma\psi$ (see the Methods section). The Hamiltonian $H$ can be decomposed into $H = D + K$, where $D$ is the diagonal matrix for the self-energy of each element (*e.g.,* wavevector $\beta$ or resonance $f$) and $K$ is the off-diagonal matrix for the interaction-energy between elements (*e.g.,* coupling $\kappa$), which represents the *network* of the system. Figure 1a shows a one-dimensional (1D) random-walk system with off-diagonal disorder[20], possessing the identical self-energy $\beta$ and disordered interactions ($[D]_p = \gamma_{00}$ and $[K]_{pq} = \kappa_{pq} = \kappa_0 + \Delta\kappa \cdot u(-1,1)$, where $\kappa_{pq}$ is the coupling between the $p^{\text{th}}$ and $q^{\text{th}}$ elements for $1 \leq p, q \leq N$, $\kappa_0$ and $\Delta\kappa$ represent the averaged and disordered coupling, respectively, and $u$ is the uniform probability density function).

Figure 1b−d presents a few eigenstates of the optical systems at different degrees of off-diagonal disorder[20] (Supplementary Note 1 for its practical realization in the mid-infrared regime). As the strength of the disorder $\Delta\kappa$ increases, Bloch eigenstates (Fig. 1b) begin to be localized (Fig. 1c), eventually exhibiting the wavelength-scale Anderson localization of exponential decay envelopes (Fig. 1d). The localization naturally results in the suppression of wave transport within the network, from the ballistic (Fig. 1e, $\alpha = 2$) to diffusive (Fig. 1f, $1 < \alpha < 2$) and even sub-diffusive (Fig. 1g, $\alpha < 1$) transports ($\alpha$: diffusion exponent[21], see the Methods section). Not restricted to eigenstate localization (Fig. 1h, $w$: modal width, see the Methods section and Supplementary Note 2) and the following suppressed wave transports (Fig. 1i), the increase of the disorder in the system also alters the spreading of its eigenvalues, linearizing the eigenband (Fig. 1j, Supplementary Note 3). In these results, a continuous transition

between the regimes of order and disorder is evident, confirming the classical relationship[2-8] between localization, transport, and disorder. Motivated by the generic form of the Hamiltonian $H = D + K$, we now demonstrate that the design of unconventional eigenstates in highly disordered potentials can be achieved by utilizing the degree of freedom on the self-energy of each element in $D$, which was neglected in Fig. 1.

Suppose that we desire to 'mold' an eigenstate $\psi$ with the *spatial* distribution of $v_m = [v_{m1}, v_{m2}, \ldots, v_{mN}]^T$ with the eigenvalue $\gamma_m$ while preserving the random-walk network of the system. For this purpose, we develop the eigen-decomposition matrix $V = [v_m, v_2, \ldots, v_N]$ using the Gram-Schmidt process, where $v_m$ and the set of column vectors $v_s = [v_{s1}, v_{s2}, \ldots, v_{sN}]^T$ ($s = 2, 3, \ldots, N$) together compose the orthonormal basis set. Due to the orthonormality ($VV^\dagger = I$), the eigenvalue equation in the $V$-reciprocal space becomes $V^\dagger HV(V^\dagger \psi) = \gamma(V^\dagger \psi)$, or $H_r \psi_r = \gamma \psi_r$, where $H_r = V^\dagger HV$ and $\psi_r = V^\dagger \psi$. The random-walk network of the system is represented by a complex matrix $K_r = V^\dagger KV$ in the $V$-reciprocal space, and then, the $V$-reciprocal Hamiltonian $H_r = V^\dagger DV + K_r$ has the following components:

$$H_r = \begin{bmatrix} \sum_p v_{mp}^* \gamma_{op} v_{mp} + K_{r11} & \sum_p v_{mp}^* \gamma_{op} v_{2p} + K_{r12} & \cdots & \sum_p v_{mp}^* \gamma_{op} v_{Np} + K_{r1N} \\ \sum_p v_{2p}^* \gamma_{op} v_{mp} + K_{r21} & \sum_p v_{2p}^* \gamma_{op} v_{2p} + K_{r22} & \cdots & \sum_p v_{2p}^* \gamma_{op} v_{Np} + K_{r2N} \\ \vdots & \vdots & \ddots & \vdots \\ \sum_p v_{Np}^* \gamma_{op} v_{mp} + K_{rN1} & \sum_p v_{Np}^* \gamma_{op} v_{2p} + K_{rN2} & \cdots & \sum_p v_{Np}^* \gamma_{op} v_{Np} + K_{rNN} \end{bmatrix}, \quad (1)$$

where $\gamma_{op} = [D]_p$ and $K_{rpq} = [K_r]_{pq}$. To design the eigenstate $\psi$ of spatial representation $v_m$, one of the reciprocal eigenstates should be $\psi_r = [1, 0, \ldots, 0]^T$. This condition is uniquely fulfilled when the first column of $H_r$ has only one nonzero component of $[H_r]_{11}$, and its eigenvalue can always be set as desired by assigning $[H_r]_{11} = \gamma_m$. We then achieve the self-energy of each element $\gamma_{op}$ deterministically as follows, from $[[H_r]_{11}, [H_r]_{12}, \ldots, [H_r]_{1N}]^T = [\gamma_m, 0, \ldots, 0]^T$:

$$\begin{bmatrix} \gamma_{o1} \\ \gamma_{o2} \\ \vdots \\ \gamma_{oN} \end{bmatrix} = \left( \begin{bmatrix} v_{m1}^* & v_{m2}^* & \cdots & v_{mN}^* \\ v_{21}^* & v_{22}^* & \cdots & v_{2N}^* \\ \vdots & \vdots & \ddots & \vdots \\ v_{N1}^* & v_{N2}^* & \cdots & v_{NN}^* \end{bmatrix} \begin{bmatrix} v_{m1} & 0 & \cdots & 0 \\ 0 & v_{m2} & \cdots & 0 \\ \vdots & \vdots & \ddots & \vdots \\ 0 & 0 & \cdots & v_{mN} \end{bmatrix} \right)^{-1} \begin{bmatrix} \gamma_m - K_{r11} \\ -K_{r21} \\ \vdots \\ -K_{rN1} \end{bmatrix}, \qquad (2)$$

or, simply, $\gamma_o = [\gamma_{o1}, \gamma_{o2}, \ldots, \gamma_{oN}]^T = [diag(v_m)]^{-1} V [\gamma_m - K_{r11}, -K_{r12}, \ldots, -K_{r1N}]^T$.

Equation (2) indicates that if there exists the inverse of the matrix $diag(v_m)$, *i.e.*, $v_{mi} \neq 0$ for all $i$, the self-energy vector $\gamma_o$ can always be found, determining the optical potential of each element from $\gamma_{op}$ (see Supplementary Fig. 1c for $\gamma_o = n$). Therefore, for 'any' networks $K$ regardless of the degree of disorder, a single eigenstate can always be molded into the desired shape of $v_m$ with the eigenvalue $\gamma_m$ by adjusting the potential of each element (Fig. 2a vs. Fig. 1a) while preserving the *network* of the system, which we herein call 'metadisorder'. The proposed metadisorder system allows for the nontrivial form of *an eigenstate* in all regimes of disordered networks $K$, for example, the globally collective eigenstate, in contrast to the case of identical elements (Fig. 1c and 1d).

Figure 2 shows examples of designer waves in 1D metadisorder systems, where the optical potential of each waveguide is calculated via Eq. (2). Compared to the highly disordered system composed of identical waveguides in Fig. 1, we provide various examples of wave systems having a designer eigenstate $v_m$: planewave (Fig. 2b), Gaussian-enveloped guided-wave with non-exponential decay (Fig. 2c), and interface (Fig. 2d) and surface (Fig. 2e) waves both with Anderson-like[2,5,6,20] exponential decay. We note that the real-valued $v_m$ corresponds to the designer ground state in the disordered eigenband (red arrows in Fig. 2f–i; see Supplementary Note 4 for designer excited states, which allow conservative waves in complex potentials).

Interestingly, our design method allows for the scattering-free planewave in highly disordered systems (Fig. 2b, $\Delta\kappa = \kappa_0$), in stark contrast to conventional disordered systems (Fig. 1g), which only lead to strong localization from random-walk scattering. The planewave eigenstate in highly disordered systems, which has a modal size equal to the overall system size, is more extended than that in finite-$N$ ordered systems with the boundary effect (Fig. 1b). Furthermore, unconventional localization forms, such as the non-Anderson Gaussian localization (Fig. 2c) or designed Anderson-like exponential localization at the interface (Fig. 2d) or surface (Fig. 2e), can also be achieved, as a general extension to the accidental emergence of classical Anderson localization (Fig. 1d). Because a potential with a globally extended eigenstate should possess the reduced random-walk scattering in the overall system, other eigenstates of similar eigenvalues also tend to have wider spatial bandwidth.

The concept of metadisorder also creates a new class of disordered potentials, which support the counterintuitive relation between eigenstate localization ($w$) and transport ($\alpha$) by imposing the designer eigenstate $v_m$ with unconventional localizations. Without loss of generality, in Fig. 3, we consider the localized designer eigenstate of the form $v_m(x) = exp[-|x|^g/(2 \cdot \sigma^g)]$ (Fig. 3a), where $g = 1$ for Anderson-like exponential localization and $g = 2$, $4$, and $6$ for the convenient examples of non-Anderson localizations. Figure 3f−n presents the localization-transport ($w$−$\alpha$) relation of 1D non-Anderson metadisorder systems compared to classical Anderson disorder (Fig. 3b) and Anderson-like metadisorder (Fig. 3c−e). Although Anderson-like metadisorders ($g = 1$) provide a similar $w$−$\alpha$ relation to that of Anderson disorder (Fig. 3c−e vs. Fig. 3b), the non-Anderson metadisorder ($g > 1$) enables more 'localized' waves yet achieves ballistic transport (*e.g.*, Fig. 3f vs. 3b, a factor of ~2 decrease for $w$ and $\alpha$ ~ 2 for $\Delta\kappa < \kappa_0 / 10$), analogous to the clustered signal transport in 'small-world' networks[1]. More interestingly, such a non-classical wave transport even enables the 'localization-induced' wave transport (increase in $\alpha$ for smaller values of $w(\Delta\kappa)$ or $\sigma$, Fig. 3h), showing the reversed relation between $w$ and $\alpha$ compared to that of conventional disordered systems[3-6,17]. This localization-induced wave transport is more apparent for metadisorder systems with larger $g$ (Fig. 3i−n for $g = 4$ and 6), allowing not only the separated control of localization and wave transport with $g$ and $\sigma$ but also the robustness of wave transports to the disorder $\Delta\kappa$, analogous to the difference between clustering and characteristic path length in small-world networks[1].

The eigenstate design in a $V$-reciprocal space allows for its extension to multidimensional problems in a straightforward manner by including all of the coupling coefficients in a multidimension in the network matrix $K$. Here, we consider two-dimensional (2D) disordered systems, obtained by the random-walk deformation of the periodic lattice (Fig. 4a). Figure 4b shows an example of disordered coupled-resonator systems from the random-walk deformation of a $17 \times 17$ square lattice (see Supplementary Note 5 for the practical realization of 2D coupled-resonator systems in the terahertz regime).

Figure 4c−h and Supplementary Movies 1−6 show collective and ordered light behaviors in highly disordered systems, which support strongly correlated phase information over the entire system. By adjusting each resonant frequency of constituent resonators following Eq. (2), we design free-form standing-wave resonances with a perfectly uniform field distribution (Fig. 4c), quadrupole phase distribution (Fig. 4d), and the designed localization (Fig. 4e) despite randomly deformed interactions between resonators. Furthermore, the introduction of complex potentials allows for one-way traveling-wave resonances, for example, by imposing the form of $exp(-ip\theta)$ on $v_m$ (Fig. 4f−h for the azimuth $\theta$). Such a 'chiral' rotation of the phase in collective resonances derives the orbital angular momentum (OAM)[22] of resonant light. Notably, although the chiral feature of light in the proposed metadisorder systems also requires complex optical potentials with gain and loss, our method involves neither PT symmetry nor periodicity[23]. *Propagating* light with nonzero OAM can also be achieved by employing waveguide-based disordered systems.

Having demonstrated collective resonance modes in highly disordered systems, finally, we present the excitation of designer eigenstates (Fig. 4i−m and Supplementary Movies 7−11) with the external coupling of conventional waveforms. When the inner connection of the system is sufficiently strong, the separation of eigenvalues (that is, free spectral range (FSR)) becomes sufficiently large to achieve wave dynamics dependent solely on a single eigenstate. Figure 4i and Supplementary Movie 7 show the wave flow through the perfectly uniform collective eigenstate over the entire system. Following the property of the eigenstate, the disordered system becomes 'invisible' for incident planewaves, prohibiting any alteration of phase and amplitude (transmission $T \sim 100\%$): zero effective index. With eigenstate-based metadisorder design, we also implement high-level functionalities with excellent throughputs, including tunable light focusing (Fig. 4j, $T \sim 96\%$), phase-conserved spatial oscillation (Fig. 4k, $T \sim 98\%$), parity converters (even to odd, Fig. 4l, $T \sim 99\%$) of real potentials, and the point-source excitation of oblique planewaves (5.6°, Fig. 4m and Supplementary Movie 11, $T \sim 97\%$) using complex potentials.

To summarize, we revealed a new class of random-walk wave systems, i.e., 'metadisorder', which can be globally collective and deliberately controlled. Exploiting metadisorders of non-Anderson-type localization, we first derive the counterintuitive relation between localization and transport, including small-world-like[1] or localization-induced transports. As demonstrated in collective wave dynamics and functionalities, our eigenstate-based approach also provides the powerful means to control wave flow while preserving or manipulating the phase information. Although we controlled only the self-energy $D$ for the Hamiltonian $H = D + K$, our method can also be easily extended to determine the 'network' $K$ of disordered self-energy distributions for the designer eigenstate. We emphasize that our method is distinct from other approaches handling the flow of waves; the supersymmetric technique[24-27] controls eigenspectra but transforms eigenstates in a fixed manner; transformation optics[28] treats a 'continuous' potential landscape, lacking the degree of freedom in interaction-energy. From their small-world inherited disorder-robustness and globally collective features, we also envisage the application of metadisorder systems to many other nontrivial physics, such as hyperuniformity[29,30], topological networks[18], or quasiparticles in disordered potentials.

**Methods**

**Analysis of random-walk discrete optical systems.** Consider the $N$-body system composed of weakly coupled optical elements. In CMT[17] or TB[19] methods, the governing equation, including self- and interaction-energy, becomes

$$\frac{d}{d\xi}\psi_p = i\gamma_{op}\psi_p + \sum_{q \neq p} i\kappa_{pq}\psi_q, \tag{3}$$

where $p = 1, 2, \ldots, N$ is the element number, $\psi_p$ is the field at the $p^{\text{th}}$ element, $\xi$ is the wave evolution axis (time $t$ for coupled resonators[13] and space $x$, $y$, or $z$ for coupled waveguides[17]), $\gamma_{op}$ is the self-energy of the $p^{\text{th}}$ element, and $\kappa_{pq}$ is the coupling coefficient between the $p^{\text{th}}$ and $q^{\text{th}}$ elements. For the steady-state solution ($\partial_\xi \rightarrow i\gamma$), Eq. (3) becomes the matrix eigenvalue problem $H\psi = \gamma\psi$, where

$$
\begin{aligned}
H &= \begin{bmatrix}
\gamma_{o1} & \kappa_{12} & \cdots & \kappa_{1N} \\
\kappa_{21} & \gamma_{o2} & \cdots & \kappa_{2N} \\
\vdots & \vdots & \ddots & \vdots \\
\kappa_{N1} & \kappa_{N2} & \cdots & \gamma_{oN}
\end{bmatrix} \\
&= D + K = \begin{bmatrix}
\gamma_{o1} & 0 & \cdots & 0 \\
0 & \gamma_{o2} & \cdots & 0 \\
\vdots & \vdots & \ddots & \vdots \\
0 & 0 & \cdots & \gamma_{oN}
\end{bmatrix} + \begin{bmatrix}
0 & \kappa_{12} & \cdots & \kappa_{1N} \\
\kappa_{21} & 0 & \cdots & \kappa_{2N} \\
\vdots & \vdots & \ddots & \vdots \\
\kappa_{N1} & \kappa_{N2} & \cdots & 0
\end{bmatrix},
\end{aligned} \tag{4}
$$

$\psi = [\psi_1, \psi_2, \ldots, \psi_N]^{\text{T}}$, $D$ is the diagonal self-energy matrix, and $K$ is the off-diagonal network matrix. The randomly coupled system can then be described by assigning random numbers to the components of the $K$ matrix. Because each element number $p$ corresponds to the physical location of the $p^{\text{th}}$ element $\mathbf{X}_p$ (*e.g.*, $p \rightarrow x_p$ for 1D and $p \rightarrow (x_p, y_p)$ for 2D problems), the obtained eigenstate $\psi$ can be re-expressed in the spatial domain $\psi = \psi(\mathbf{X})$.

**Calculation of the diffusion exponent $\alpha$.** Consider the 1-dimensional system of Fig. 1 ($\xi = y$), which has eigenstates $\psi_k(x)$ and corresponding eigenvalues $\gamma_k$ ($k = 1, 2, \ldots, N$; $\gamma_k$ is the effective wavevector of $\psi_k$). For the incidence of $\varphi_i(x) = \Sigma a_k \cdot \psi_k$, the propagating field can be obtained via the TMM as $\varphi(x,y) = \Sigma a_k \cdot \psi_k \cdot exp(i\gamma_k \cdot y)$. To analyze the transporting feature of the system without boundary effects, the incident wave is excited at the center waveguide ($x = x_m$, where $m = (N + 1) / 2$ for odd $N$); we can then calculate the spatially varying mean-square displacement (MSD)[21] $M(y)$ as follows:

$$M(y) = \langle x^2 \rangle = \frac{\sum_p (x_p - x_m)^2 \cdot |\varphi(x_p, y)|^2}{\sum_p |\varphi(x_p, y)|^2} . \tag{5}$$

When the MSD $M(y)$ is fitted for $y$ exponentially as $M(y) \sim c_\alpha \cdot y^\alpha$, we achieve the diffusion exponent $\alpha$: $\alpha = 2$ for ballistic transport and $\alpha = 1$ for diffusive transport[21]. The calculated results are shown in Figs 1e−g, 1i and 3. Note that $m$ does not have to be the center waveguide precisely when $N$ is sufficiently large and thus the boundary effect can be neglected.

**Calculation of modal size.** For the 1-dimensional system ($\xi = y$) with $\psi_k(x)$ and $\gamma_k$, the modal size for each eigenstate is defined as[5]

$$w_k = \frac{\left[ \sum_{p=1}^{N} |\psi_k(x_p)|^2 \cdot \Delta x_p \right]^2}{\sum_{p=1}^{N} |\psi_k(x_p)|^4 \cdot \Delta x_p} , \tag{6}$$

where $\Delta x_p$ is the size of the $p^{\text{th}}$ element, obtained from the distance between waveguides for each value of the coupling coefficient (see Supplementary Note 1). See also Supplementary Note 2 for eigenstate-dependent localizations.

**Acknowledgments**

The authors thank S. Torquato for his encouragement of our research on disordered systems. This work was supported by the National Research Foundation of Korea through the Global Frontier Program (GFP) NRF-2014M3A6B3063708 and the Global Research Laboratory (GRL) Program K20815000003, which are all funded by the Ministry of Science, ICT & Future Planning of the Korean government.

**Author Contributions**

S.Y. conceived the presented idea. S.Y. developed the theory and performed the computations. X.P. contributed to the analysis of wave transports. X.P. and J.H. verified the analytical methods. N.P. encouraged S.Y. to investigate the inverse design and small-world nature of eigensystems while supervising the findings of this work. All authors discussed the results and contributed to the final manuscript.

**Competing Interests Statement**

The authors declare that they have no competing financial interests.

**References**

1. Watts, D. J. & Strogatz, S. H. Collective dynamics of 'small-world' networks. *Nature* **393**, 440-442 (1998).

2. Anderson, P. W. Absence of diffusion in certain random lattices. *Phys. Rev.* **109**, 1492 (1958).

3. Wiersma, D. S. Disordered photonics. *Nature Photon.* **7**, 188-196 (2013).

4. Wiersma, D. S., Bartolini, P., Lagendijk, A. & Righini, R. Localization of light in a disordered medium. *Nature* **390**, 671-673 (1997).

5. Schwartz, T., Bartal, G., Fishman, S. & Segev, M. Transport and Anderson localization in disordered two-dimensional photonic lattices. *Nature* **446**, 52-55 (2007).

6. Lahini, Y. *et al.* Anderson localization and nonlinearity in one-dimensional disordered photonic lattices. *Phys. Rev. Lett.* **100**, 013906 (2008).

7. Papasimakis, N., Fedotov, V. A., Fu, Y. H., Tsai, D. P. & Zheludev, N. I. Coherent and incoherent metamaterials and order-disorder transitions. *Phys. Rev. B* **80**, 041102 (2009).

8. Poddubny, A. N., Rybin, M. V., Limonov, M. F. & Kivshar, Y. S. Fano interference governs wave transport in disordered systems. *Nature Comm.* **3**, 914 (2012).

9. Abe, S. & Suzuki, N. Small-world structure of earthquake network. *Physica A: Statistical Mechanics and its Applications* **337**, 357-362 (2004).

10. Li, S. *et al.* A map of the interactome network of the metazoan C. elegans. *Science* **303**, 540-543 (2004).

11. Bullmore, E. T. & Bassett, D. S. Brain graphs: graphical models of the human brain connectome. *Annual review of clinical psychology* **7**, 113-140 (2011).

12.     Christodoulides, D. N., Lederer, F. & Silberberg, Y. Discretizing light behaviour in linear and nonlinear waveguide lattices. *Nature* **424**, 817-823 (2003).

13.     Joannopoulos, J. D., Johnson, S. G., Winn, J. N. & Meade, R. D. *Photonic crystals: molding the flow of light*.   (Princeton university press, 2011).

14.     Leonetti, M., Karbasi, S., Mafi, A. & Conti, C. Light focusing in the Anderson regime. *Nature Comm.* **5** (2014).

15.     Mosk, A. P., Lagendijk, A., Lerosey, G. & Fink, M. Controlling waves in space and time for imaging and focusing in complex media. *Nature Photon.* **6**, 283-292 (2012).

16.     Park, J.-H. *et al.* Subwavelength light focusing using random nanoparticles. *Nature Photon.* **7**, 454-458 (2013).

17.     Garanovich, I. L., Longhi, S., Sukhorukov, A. A. & Kivshar, Y. S. Light propagation and localization in modulated photonic lattices and waveguides. *Phys. Rep.* **518**, 1-79 (2012).

18.     Liang, G. & Chong, Y. Optical resonator analog of a two-dimensional topological insulator. *Phys. Rev. Lett.* **110**, 203904 (2013).

19.     Lidorikis, E., Sigalas, M., Economou, E. N. & Soukoulis, C. Tight-binding parametrization for photonic band gap materials. *Phys. Rev. Lett.* **81**, 1405 (1998).

20.     Martin, L. *et al.* Anderson localization in optical waveguide arrays with off-diagonal coupling disorder. *Opt. Express* **19**, 13636-13646 (2011).

21.     Hahn, K., Kärger, J. & Kukla, V. Single-file diffusion observation. *Phys. Rev. Lett.* **76**, 2762 (1996).

22.     Molina-Terriza, G., Torres, J. P. & Torner, L. Twisted photons. *Nature Phys.* **3**, 305-310 (2007).

23.    Yu, S., Park, H. S., Piao, X., Min, B. & Park, N. Chiral interactions of light induced by low-dimensional dynamics in complex potentials. *arXiv preprint arXiv:1409.0180* (2014).

24.    Yu, S., Piao, X., Hong, J. & Park, N. Bloch-like waves in random-walk potentials based on supersymmetry. *Nature Comm.* **6**, 8269 (2015).

25.    Miri, M.-A., Heinrich, M., El-Ganainy, R. & Christodoulides, D. N. Supersymmetric optical structures. *Phys. Rev. Lett.* **110**, 233902 (2013).

26.    Longhi, S. Supersymmetric transparent optical intersections. *Opt. Lett.* **40**, 463-466 (2015).

27.    Miri, M.-A., Heinrich, M. & Christodoulides, D. N. Supersymmetry-generated complex optical potentials with real spectra. *Phys. Rev. A* **87**, 043819 (2013).

28.    Pendry, J. B., Schurig, D. & Smith, D. R. Controlling electromagnetic fields. *Science* **312**, 1780-1782 (2006).

29.    Torquato, S. & Stillinger, F. H. Local density fluctuations, hyperuniformity, and order metrics. *Phys. Rev. E* **68**, doi:10.1103/PhysRevE.68.041113 (2003).

30.    Torquato, S., Zhang, G. & Stillinger, F. Ensemble Theory for Stealthy Hyperuniform Disordered Ground States. *Phys. Rev. X* **5**, 021020 (2015).

**Figure Legends**

**Figure 1. Effect of disorder in optical systems.** (**a**) A schematic of a random-walk optical system composed of coupled waveguides, analogous to the randomly coupled pendulums with identical oscillating features. The oscillation of each pendulum describes the phase evolution during the propagation. (**b**−**d**) The first three eigenstates for (**b**) ordered, (**c**) weakly disordered, and (**d**) Anderson potentials, calculated by using CMT in the Methods section. The potential $n$ denotes the effective waveguide index of a single waveguide. Corresponding wave transports calculated by using the transfer-matrix method (TMM) are shown in (**e**−**g**), respectively (see the Methods section). The variations of (**h**) modal size $w$ and (**i**) diffusion exponent $\alpha$ are shown as a function of the disorder $\Delta\kappa$. (**j**) Eigenbands ($n_{\text{eff}}$) for the disorder in (**b**, black dotted line) and (**d**, green points and line). The points in (**h**−**j**) represent each statistical ensemble, and solid lines are the averages of 200 ensembles. The green dotted line in (**i**) denotes the diffusion state ($\alpha = 1$). $\gamma_{o0} = 1.6 \cdot k_0$, $\kappa_0 = 0.01 \cdot k_0$, and $N = 51$ for (**b**−**j**), where $k_0 = 2\pi / \lambda_0$ is the free-space wavenumber. The practical waveguide design and the distance between waveguides in the mid-infrared regime ($\lambda_0 = 3$ μm) are calculated in Supplementary Note 1 using COMSOL Multiphysics.

**Figure 2. 1D metadisorder systems.** (**a**) A schematic of a 1D metadisorder system composed of coupled waveguides, analogous to the randomly coupled pendulums with different self-oscillating features, such as oscillating period (rod length) and gain or loss parameters (color). Each waveguide has different real parts of self-energy due to changing the width of the waveguide (Supplementary Note 1). The colors of the waveguides represent the imaginary part of self-energy: gain and loss (treated in Supplementary Note 4). (**b**−**e**) Designed eigenstates and optical potentials, eigenstate propagations, and (**f**−**i**) eigenvalues ($n_{\text{eff}}$) of 1D metadisorder systems are calculated by using the CMT for (**b**, **f**) planewave ($v_{\text{m}}(x) = 1$), (**c**, **g**) Gaussian wave ($v_{\text{m}}(x) = exp[-x^2/(2 \cdot \sigma^2)]$), (**d**, **h**) interface wave ($v_{\text{m}}(x) = exp[-|x|/(2 \cdot \sigma)]$), and (**e**, **i**) surface wave ($v_{\text{m}}(x) = exp[-|x-x_{\text{L}}|/(2 \cdot \sigma)]$) eigenstates, where the spatial bandwidth $\sigma = L_{\text{st}} / 16$ in (**c**−**e**, **g**−**i**), the left boundary $x_{\text{L}} = -L_{\text{st}} / 2$, and $L_{\text{st}}$ is the overall potential length. $\Delta\kappa = \kappa_0$ in (**b**−**e**) for the extreme degree of disorder. Blue symbols represent $\Delta\kappa = \kappa_0$, green symbols represent $\Delta\kappa = 0.53\cdot\kappa_0$ and black dotted lines represent $\Delta\kappa = 0$ in (**f–i**). All of the other parameters are the same as those in Fig. 1 based on Supplementary Note 1.

**Figure 3. Non-Anderson metadisorder systems with counterintuitive wave transports.** (**a**) Shapes of designer eigenstates $v_m(x) = exp[-|x|^g/(2\cdot\sigma^g)]$ with different $g$. Eigenstate localization ($w$) and wave transport ($\alpha$) for (**b**) Anderson disorder with identical elements, (**c–e**) $g = 1$ Anderson-like metadisorders, and (**f–h**) $g = 2$, (**i–k**) $g = 4$, and (**l–n**) $g = 6$ non-Anderson metadisorders. In metadisorder systems, the bandwidths of designer eigenstates are $\sigma = L_{st} / 16$ in (**c, f, i, l**), $\sigma = L_{st} / 24$ in (**d, g, j, m**), and $\sigma = L_{st} / 32$ in (**e, h, k, n**). Error bars in (**b–n**) denote the standard deviation of 200 ensembles. All of the other parameters are the same as those in Fig. 1 based on Supplementary Note 1.

**Figure 4. 2D metadisorder systems.** Schematics of (**a**) a coupled-resonator lattice with identical elements (left) and its metadisorder-transformed structure (right). The self-energy of each element in the metadisorder system is adjusted by controlling the size of the resonator or using gain or loss materials. Nearest-neighbor (yellow arrows) and next-nearest-neighbor (green arrows) couplings are presented. (**b–m**) 2D metadisorder systems. The practical weak-coupling design for the CMT analysis are calculated in Supplementary Note 5, using COMSOL Multiphysics, for the terahertz regime ($\lambda_0 = c / f_0 = 265.3 \ \mu m$). For the periodicity of $a_x = a_y = 0.16\cdot\lambda_0$, the position of each resonator is randomly shifted by $\Delta_x = \Delta_y = 0.03\cdot\lambda_0\cdot u(-1, 1)$ in (**b–m**), and the self-energy of each element is adjusted by $f_0\cdot(1 + \Delta_f)$ following the metadisorder design from Eq. (2). (**b**) A sample of obtained resonator distribution for the $17 \times 17$ lattice in (**c–h**). (**c–e**) Standing-wave collective resonances using real potentials for (**c**) uniform, (**d**) quadrupole, and (**e**) exponentially localized distributions. (**f–h**) Traveling-wave chiral collective resonances, using complex potentials, for (**f**) dipole, (**g**) quadrupole, and (**h**) octopole distributions. (**i–m**) Metadisorder-based functionalities in the $17 \times 17$ lattice for (**i**) invisible disorder, (**j**) steered focusing, (**k**) spatial oscillation, (**l**) parity-converted beam splitter, and (**m**) the point-source excitation of oblique planewaves. The designed resonance is set to $\gamma_m = f_0$ for all cases. See Supplementary Movies 1–11 for the temporal dynamics of each case in (**c–m**).

**Figures**

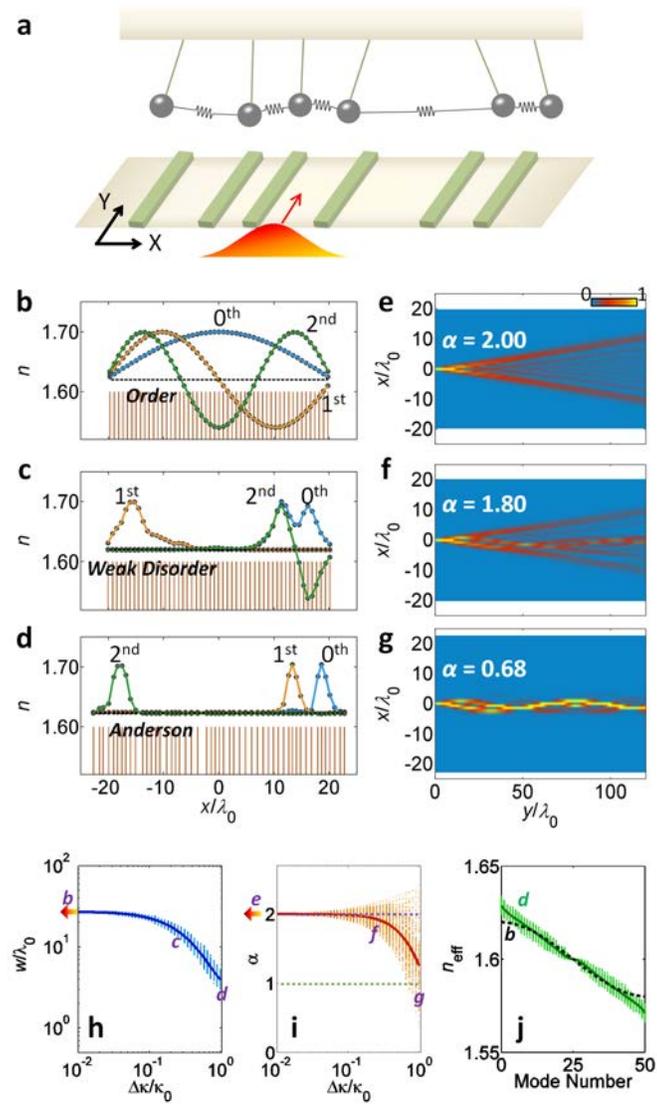

**Yu, Figure 1**

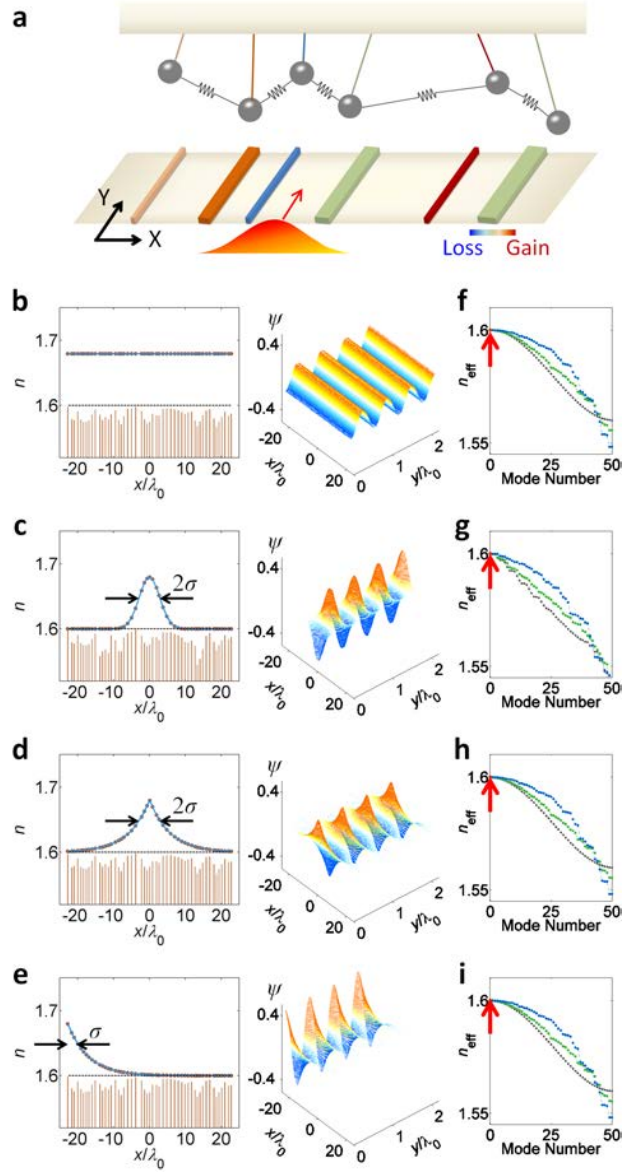

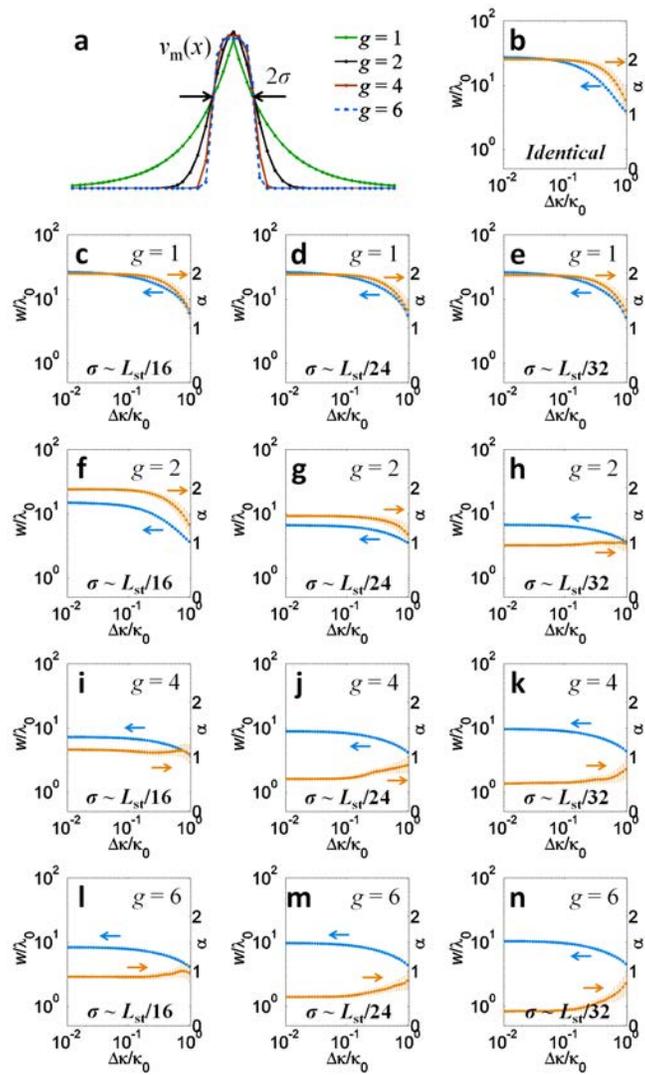

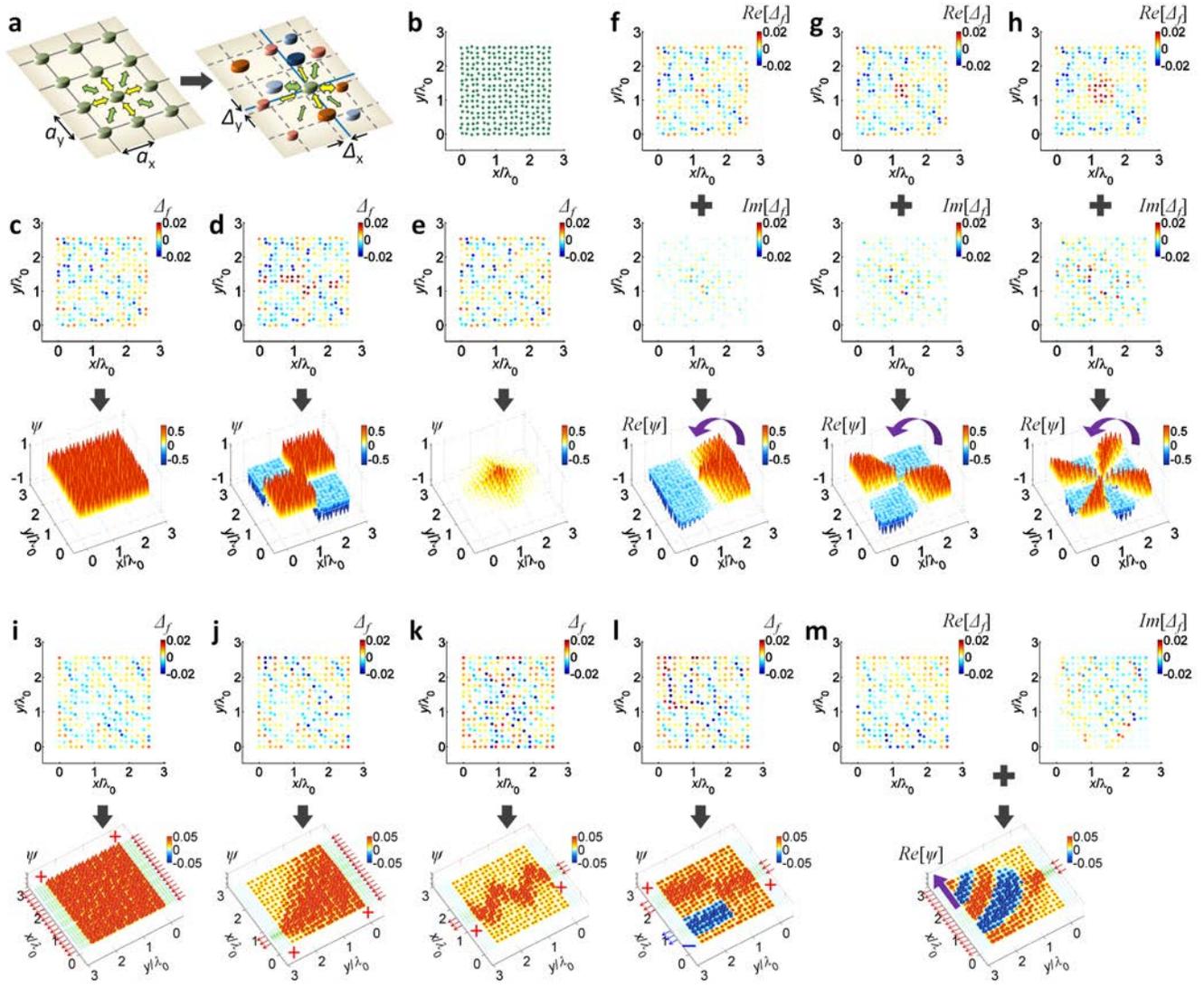

Yu, Figure 4

**Supplementary Note 1. Design of waveguide-based 1D random-walk networks**

Supplementary Figure 1a shows the schematic of coupled waveguides as a basic element for 1D random-walk networks. We assume silicon nitride[1] waveguides ($Si_3N_4$, refractive index $n = 2.43$) embedded in a silicon dioxide[1] matrix ($SiO_2$, $n = 1.42$), operating in the mid-infrared regime (free-space wavelength of $\lambda_0 = 3$ µm). The $E_{11}{}^x$ waveguide mode[2] is used as a fundamental mode (Supplementary Fig. 1b). The self-energy (waveguide index $n = \beta/k_0$) can be controlled by the waveguide width $w$, deriving the quasi-linear relation between $n_{eff} = 1.53$ to $1.66$ and $w = 800$ to $1100$ nm (Supplementary Fig. 1c), which sufficiently covers the entire parameter space used in Figs 2 and 3 in the main manuscript.

For the weak modulation of the self-energy ($n_{eff} = 1.53$ to $1.66$), the strength of the interaction-energy (coupling coefficient $\kappa$) can be manipulated independently from the change of self-energy due to the negligible magnitude of the overlap between the evanescent field (for $\kappa$) and the perturbed field (for the variation of $n_{eff}$)[3,4]. To illustrate this minor effect of structural perturbations ($w_{1,2}$) on the coupling $\kappa$, we scan the variation of $\kappa$ for different combinations of $w_1$ and $w_2$ (Supplementary Fig. 1d−f). Supplementary Fig. 1g shows the error bar plot of $\kappa$ as a function of the waveguide separation $d$ for different pairs of $w_1$ and $w_2$ ($800$ nm $\leq w_{1,2} \leq 1100$ nm). As shown, $\kappa$ is primarily determined by $d$, and the exponential function $\kappa/k_0 \sim c_1 \cdot exp(-c_2 \cdot d/\lambda_0)$ provides an excellent fit[5] to $\kappa$ in the weak-coupling regime (here, $\kappa/\beta < 1/50$), where $c_1 = 0.462$ and $c_2 = 4.85$.

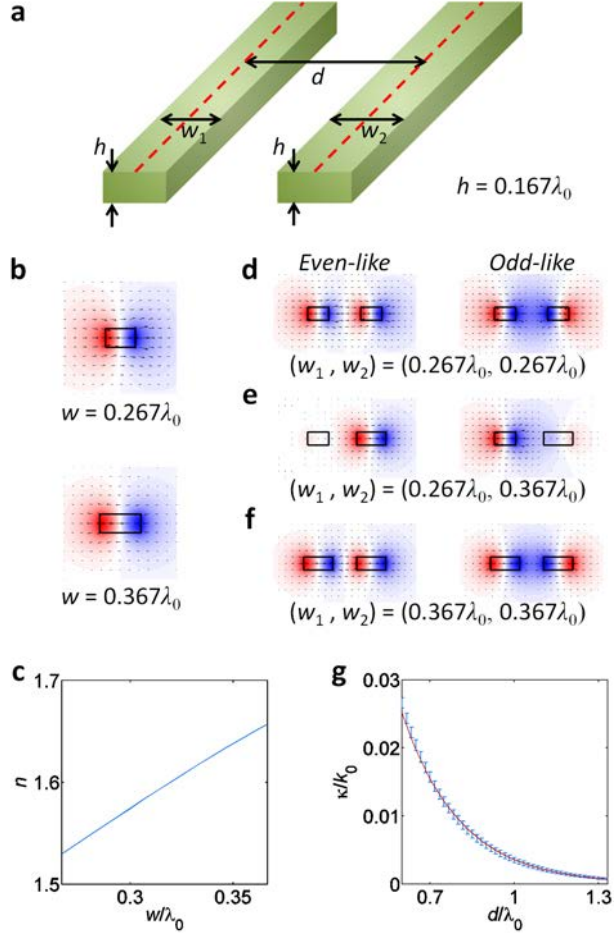

**Supplementary Figure 1. The design of waveguide-based elements for 1D random-walk networks.** (**a**) A schematic of coupled waveguides. Green regions denote $Si_3N_4$ waveguides, embedded in a $SiO_2$ matrix. The height $h$ = 500 nm. (**b**) Longitudinal electric field profiles of the fundamental eigenstate in a single waveguide for $w$ = 800 nm (up) and 1100 nm (down). (**c**) Waveguide index $n$ of a single waveguide as a function of the waveguide width $w$. (**d–f**) Longitudinal electric field profiles of coupled eigenstates for (**d**) $(w_1, w_2)$ = (800 nm, 800 nm), (**e**) $(w_1, w_2)$ = (800 nm, 1100 nm), and (**f**) $(w_1, w_2)$ = (1100 nm, 1100 nm) at the waveguide separation $d$ = 2 μm. Arrows in (**b, d–f**) denote the transverse electric field. (**g**) Normalized coupling coefficient $\kappa/k_0$ as a function of $d$. Error bars denote the standard deviation for the set of 256 combinations of $(w_1, w_2)$ (800 nm $\leq w_{1,2} \leq$ 1100 nm with 20 nm intervals). Red line is the exponential fitting curve of $\kappa/k_0 \sim c_1 \cdot exp(-c_2 \cdot d/\lambda_0)$. All of the results are obtained using COMSOL Multiphysics.

**Supplementary Note 2. Eigenstate-dependent localization**

Supplementary Figure 2 shows the localization property of each eigenstate. As observed, the modes near eigenband edges (modal numbers 0 ~ 11, 39 ~ 50, Supplementary Fig. 2a, c) have smaller modal widths than those of the modes near the center of the eigenband[6-8] (modal numbers 12 ~ 38, Supplementary Fig. 2b). Such a stronger localization near band edges corresponds to smaller transverse group velocity, hindering wave transport.

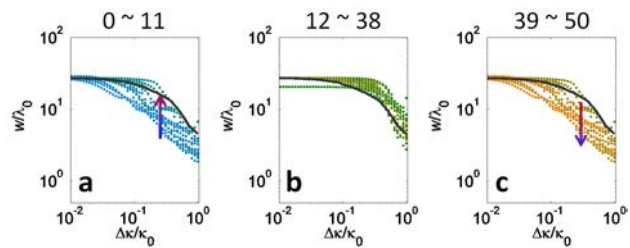

**Supplementary Figure 2. The localization of each eigenstate** for modal numbers (**a**) 0−11, (**b**) 12−38, and (**c**) 39−50. The points represent each eigenstate of a random-walk system ($N = 51$), and the black line denotes the averaged modal width. The arrows in (**a, c**) show the variation of the modal width for increasing modal number. All of the parameters are the same as those in Fig. 1 in the main manuscript.

**Supplementary Note 3. The variation of the eigenband by disorders**

Supplementary Figure 3 shows the variation of the eigenband by increasing the degree of disorder. As $\Delta\kappa$ increases, the conventional cosine-form eigenband of the periodic potential becomes linearized (Supplementary Figure 3a−f). This linearization, which corresponds to the uniformly separated eigenspectrum, originates from the broadened spectral distribution of the coupling coefficient.

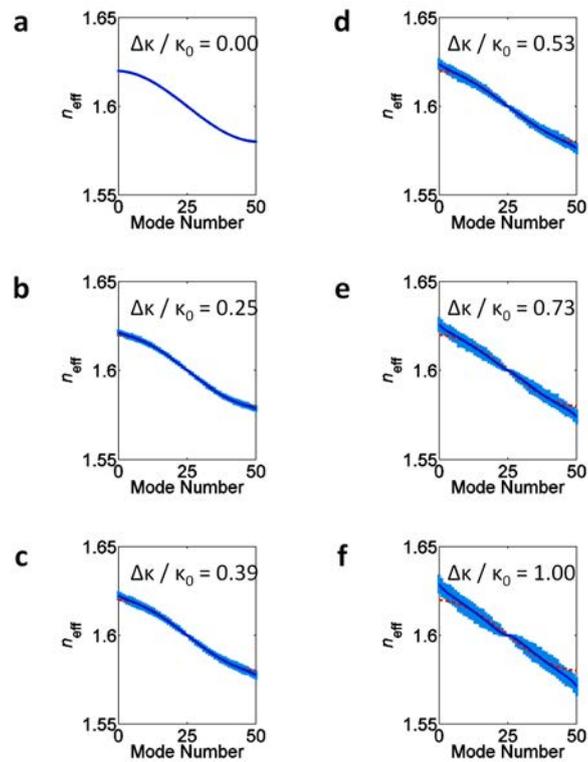

**Supplementary Figure 3. The variation of the eigenband for different degrees of disorder $\Delta\kappa$.** The points represent each eigenstate of a random-walk system, and blue lines denote the averaged eigenband. All of the parameters are the same as those in Fig. 1 in the main manuscript.

**Supplementary Note 4. Designer excited states in 1D metadisorder systems**

To obtain designer excited states, their spatial form $v_m$ should derive the real part of the transverse wavevector ($x$-axis in Figs 1 and 2 in the main manuscript), corresponding to oblique planewaves with the complex-valued $v_m$. Supplementary Figure 4a–d shows the form of the designed excited eigenstate and optical potentials. Due to the complex-valued matrix $V = [v_m, v_2, ..., v_N]$, optical potentials from $\gamma_o$ (Eq. (2) in the main manuscript) also have complex values.

From Eq. (2), the distribution of real and imaginary potentials is determined by the speed of phase evolution along the $x$-axis (or the magnitude of the tilted angle, Supplementary Fig. 4a–c). The angle can be reversed by assigning the complex-conjugated designer potential $\gamma_o{}^*$ (Supplementary Figs 4b vs 4d). Furthermore, with the freedom in the selection of the eigenvalue in Eq. (2), we assign real $\gamma_m$ to realize conservative, perfect planewaves of oblique propagation (Supplementary Fig. 4e–h, maximum 13° with feasible material parameters[9-12]). As expected, the oblique propagation corresponds to an excited state in the eigenband (Supplementary Fig. 4i–l), in contrast to the case of normal propagations (Fig. 2 in the main manuscript), which are supported by ground states. In contrast to other methods based on parity-time symmetry[13,14] or supersymmetric transformation[15] for real eigenvalues in complex potentials, it is emphasized that our method enables the molding of the 'desired' eigenstate.

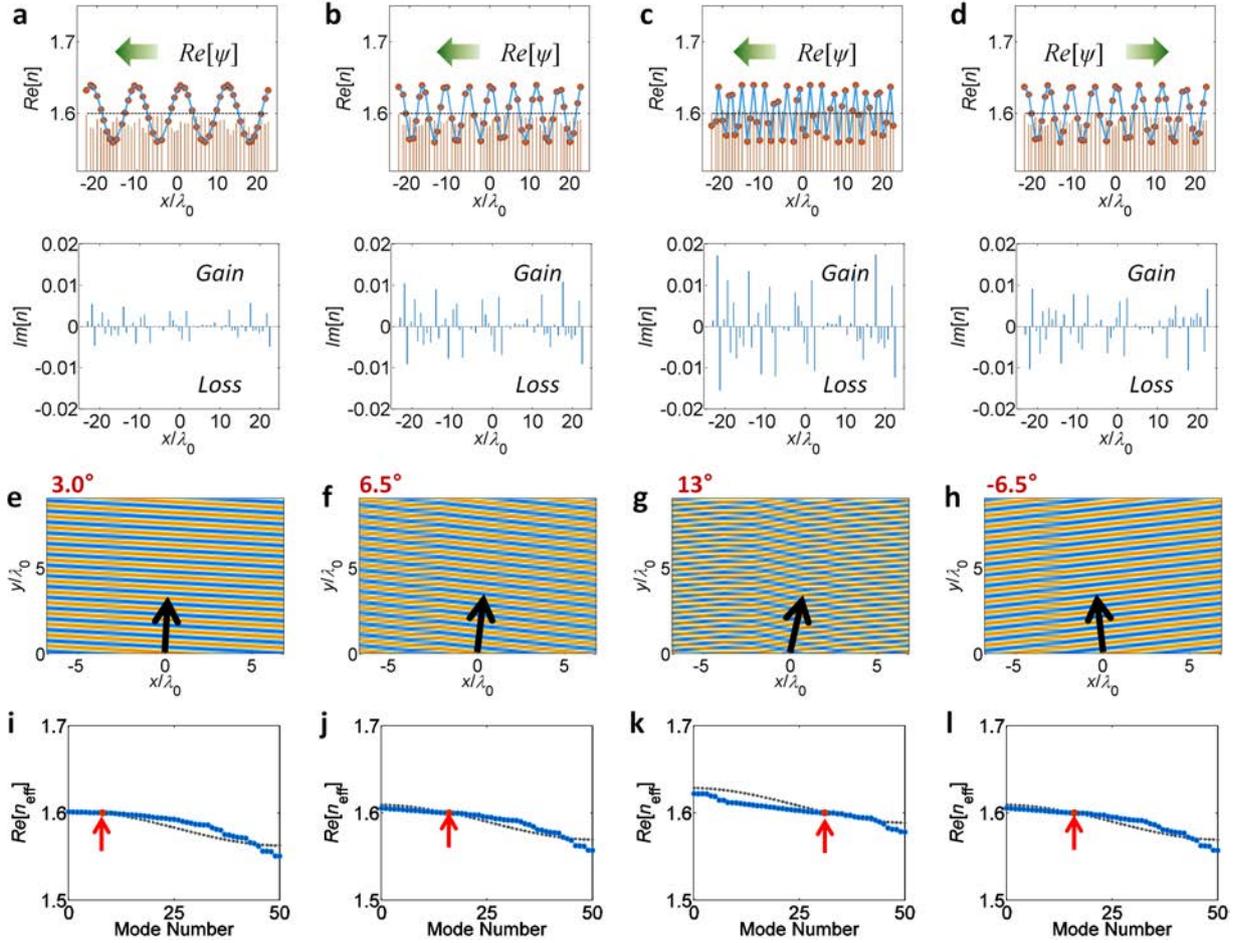

**Supplementary Figure 4. 1D metadisorder systems with complex potentials.** (**a−d**) Designed eigenstates, real ($Re[n]$) and imaginary ($Im[n]$) optical potentials, (**e−h**) eigenstate propagations, and (**i−l**) the real part of eigenvalues ($n_{eff}$) are shown for the $v_m(x)$ of (**a, e, i**) $exp(-i \cdot 8\pi \cdot x/L_{st})$, (**b, f, j**) $exp(-i \cdot 16\pi \cdot x/L_{st})$, (**c, g, k**) $exp(-i \cdot 32\pi \cdot x/L_{st})$, and (**d, h, l**) $exp(i \cdot 16\pi \cdot x/L_{st})$, where $L_{st}$ is the overall potential length. Green arrows represent the direction of the phase evolution in (**a−d**), black arrows represent the wave propagation direction in (**e−h**), and red arrows indicate the designed eigenstates in (**i−l**). $\Delta\kappa = \kappa_0$ in (**a−h**). Blue symbols represent $\Delta\kappa = \kappa_0$ and black dotted lines represent $\Delta\kappa = 0$ in (**i−l**). $\gamma_{o0} = 1.6 \cdot k_0$, $\kappa_0 = 0.01 \cdot k_0$, and $N = 51$. All of the parameters are based on the design in the mid-infrared regime ($\lambda_0 = 3$ μm), shown in Supplementary Note 1, using COMSOL Multiphysics.

**Supplementary Note 5. Design of resonator-based 2D random-walk optical networks**

Supplementary Figure 5a shows the schematic of coupled resonators as a basic element for 2D random-walk networks. We assume titanium oxide[16,17] waveguides (TiO$_2$, refractive index $n = 10$) embedded in an indium antimonide crystalline compound (InSb, $n = 0.3619 + 0.107i$), operating in the terahertz regime (target free-space wavelength of $\lambda_0 = c / f_0 = 265.3$ μm). For simplicity, we design random-walk networks that have the coupling coefficient dependent only on the resonator separation $d$. For this purpose, the transverse magnetic (TM) 'monopole' resonance is used as a fundamental mode (Supplementary Fig. 5b). The self-energy (resonant frequency $f_r$) can be controlled by the resonator radius $R$, deriving the quasi-linear relation between $f_r / f_0 = 1.11$ to $0.93$ and $R = 10.5$ to $12.5$ μm (Supplementary Fig. 5c), which sufficiently covers the entire parameter space used in Fig. 4 in the main manuscript.

Similar to the case of waveguides in Supplementary Note 1, the strength of the interaction-energy (coupling coefficient $\kappa$) can be manipulated independently from the change of self-energy. We scan the variation of $\kappa$ for different combinations of $R_1$ and $R_2$ (Supplementary Fig. 5d−f). Supplementary Fig. 5g shows the error bar plot of $\kappa$ as a function of the resonator separation $d$ for different pairs of $R_1$ and $R_2$ ($10.5$ μm $\leq R_{1,2} \leq 12.5$ μm). As shown, $\kappa$ is primarily determined by $d$, and the exponential function $\kappa/f_0 \sim c_1 \cdot exp(-c_2 \cdot d/\lambda_0)$ provides an excellent fit[5] to $\kappa$ in the weak-coupling regime (here, $\kappa/f_r < 1/40$), where $c_1 = 0.959$ and $c_2 = 36.0$.

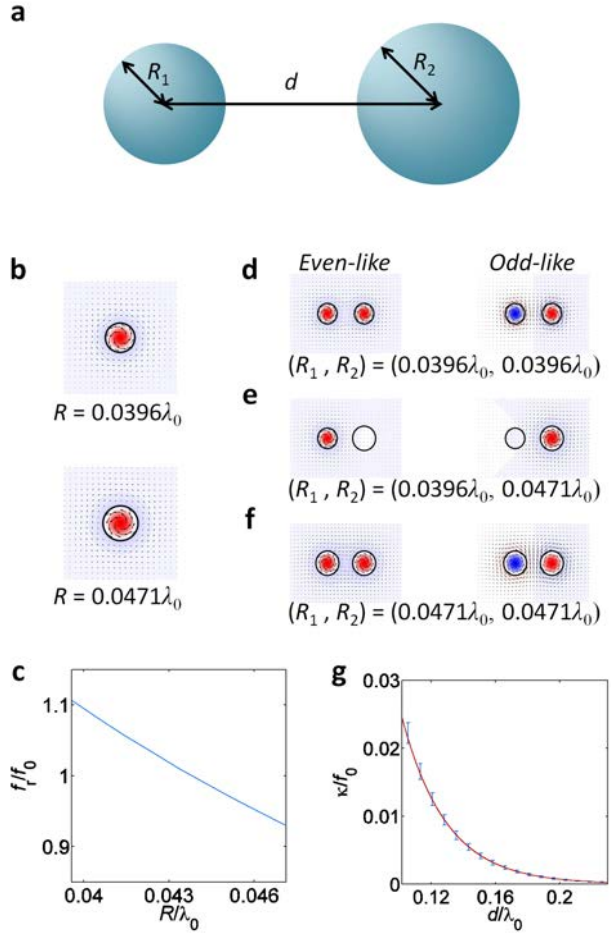

**Supplementary Figure 5. The design of resonator-based elements for 2D random-walk networks.** (**a**) A schematic of coupled resonators. Emerald regions denote $TiO_2$ resonators embedded in an InSb crystalline compound. (**b**) Longitudinal magnetic field profiles of the fundamental eigenstate in a single resonator for $w = 10.5$ μm (up) and 12.5 μm (down). (**c**) Resonant frequency $f_r$ of a single resonator, normalized by $f_0$, as a function of the resonator radius $R/\lambda_0$, where $\lambda_0 = 265.3$ μm and $f_0 = c/\lambda_0$. (**d–f**) Longitudinal magnetic field profiles of coupled eigenstates for (**d**) $(R_1, R_2) = (10.5$ μm, 10.5 μm), (**e**) $(R_1, R_2) = (10.5$ μm, 12.5 μm), and (**f**) $(R_1, R_2) = (12.5$ μm, 12.5 μm) at the resonator separation $d = 40$ μm. Arrows in (**b, d–f**) denote the transverse electric field. (**g**) Normalized coupling coefficient $\kappa/f_0$ as a function of $d$. Error bars denote the standard deviation for the set of 81 combinations of $(R_1, R_2)$ (10.5 μm $\leq w_{1,2} \leq 12.5$ μm with 250 nm intervals). Red line is the exponential fitting curve of $\kappa/f_0 \sim c_1 \cdot exp(-c_2 \cdot d/\lambda_0)$. All of the results are obtained using COMSOL Multiphysics.

# References

1. Kischkat, J. *et al.* Mid-infrared optical properties of thin films of aluminum oxide, titanium dioxide, silicon dioxide, aluminum nitride, and silicon nitride. *Appl. Opt.* **51**, 6789-6798 (2012).

2. Okamoto, K. *Fundamentals of optical waveguides.* (Academic press, 2010).

3. Joannopoulos, J. D., Johnson, S. G., Winn, J. N. & Meade, R. D. *Photonic crystals: molding the flow of light.* (Princeton university press, 2011).

4. Haus, H. A. Waves and Fields in Optoelectronics (Prentice-Hall series in solid state physical electronics). (1984).

5. Szameit, A., Dreisow, F., Pertsch, T., Nolte, S. & Tünnermann, A. Control of directional evanescent coupling in fs laser written waveguides. *Opt. Express* **15**, 1579-1587 (2007).

6. de Moura, F. A. & Lyra, M. L. Delocalization in the 1D Anderson model with long-range correlated disorder. *Phys. Rev. Lett.* **81**, 3735 (1998).

7. Theodorou, G. & Cohen, M. H. Extended states in a one-demensional system with off-diagonal disorder. *Phys. Rev. B* **13**, 4597 (1976).

8. Lahini, Y. *et al.* Anderson localization and nonlinearity in one-dimensional disordered photonic lattices. *Phys. Rev. Lett.* **100**, 013906 (2008).

9. Saitoh, T. & Mukai, T. 1.5 μm GaInAsP traveling-wave semiconductor laser amplifier. *Quantum Electronics, IEEE Journal of* **23**, 1010-1020 (1987).

10. Laming, R., Zervas, M. N. & Payne, D. N. Erbium-doped fiber amplifier with 54 dB gain and 3.1 dB noise figures. *Photonics Technology Letters, IEEE* **4**, 1345-1347 (1992).

11. Bakonyi, Z. *et al.* High-gain quantum-dot semiconductor optical amplifier for 1300 nm. *Quantum Electronics, IEEE Journal of* **39**, 1409-1414 (2003).

12.    De Leon, I. & Berini, P. Amplification of long-range surface plasmons by a dipolar gain medium. *Nature Photon.* **4**, 382-387, doi:10.1038/nphoton.2010.37 (2010).

13.    Rüter, C. E. *et al.* Observation of parity–time symmetry in optics. *Nature Phys.* **6**, 192-195 (2010).

14.    Bender, C. M. & Boettcher, S. Real spectra in non-Hermitian Hamiltonians having PT symmetry. *Phys. Rev. Lett.* **80**, 5243 (1998).

15.    Miri, M.-A., Heinrich, M. & Christodoulides, D. N. Supersymmetry-generated complex optical potentials with real spectra. *Phys. Rev. A* **87**, 043819 (2013).

16.    Matsumoto, N. *et al.* Analysis of dielectric response of TiO2 in terahertz frequency region by general harmonic oscillator model. *Jpn. J. Appl. Phys.* **47**, 7725 (2008).

17.    Berdel, K., Rivas, J. G., Bolívar, P. H., De Maagt, P. & Kurz, H. Temperature dependence of the permittivity and loss tangent of high-permittivity materials at terahertz frequencies. *Microwave Theory and Techniques, IEEE Transactions on* **53**, 1266-1271 (2005).